\documentstyle[12pt]{article}
\begin{document}

\noindent Stockholm\\
October 2006\\

\vspace{10mm}

\begin{center}

{\Large THREE WAYS TO LOOK} 

\vspace{8mm}

{\Large AT}

\vspace{8mm}

{\Large MUTUALLY UNBIASED BASES}\footnote{Talk at the 
V\"axj\"o conference on Foundations of Probability and Physics, 
June 2006.} 

\vspace{15mm}

{\large Ingemar Bengtsson}\footnote{Email address: ingemar@physto.se. 
Supported by VR.}

\vspace{10mm}

{\sl Stockholm University, Alba Nova\\
Fysikum\\
S-106 91 Stockholm, Sweden}

\vspace{15mm}

{\bf Abstract}

\end{center}

\vspace{5mm}

\noindent This is a review of the problem of Mutually Unbiased Bases in finite 
dimensional Hilbert spaces, real and complex. Also a 
geometric measure of ``mubness'' is introduced, and applied to some 
explicit calculations in six dimensions (partly done by Bj\"orck and by 
Grassl). Although this does not yet solve any problem, some appealing 
structures emerge.

\newpage

{\bf 1. The problem}

\vspace{5mm}

\noindent A formula that has been of central importance in many discussions 
about the foundations of quantum mechanics is 

\begin{equation} |\langle x|p\rangle |^2 = \mbox{constant} \ . \end{equation} 

\noindent It expresses the complementarity of position and momentum. If we 
know everything about position, we know nothing of momentum. In Hilbert 
spaces of finite dimension $N$, the analogous equation would concern two orthonormal 
bases ${|e_a\rangle }$ and ${|f_a\rangle }$ such that  

\begin{equation} |\langle e_a|f_b\rangle |^2 = \mbox{constant} = 
\frac{1}{N} \ , \hspace{5mm} 0 \leq a, b \leq N - 1 \ . \label{1} \end{equation}

\noindent The important thing is that the right hand side is independent 
of $a$ and $b$. If such bases can be found, they are said to be mutually 
unbiased bases, or MUBs for short. The name emphasises that the information 
obtained in a projective measurement associated to one of the bases is 
completely unrelated to the information obtained from a projective 
measurement associated to the other.  
The question is: how many MUBs can one introduce in a given Hilbert space? 
If $N$ is a power of a prime number then one 
can find $N + 1$ MUBs \cite{Ivanovic} \cite{Wootters2}, but for other 
values of $N$ the number might be smaller. On the other hand it might not: 
I simply do not know. 

Why should you care about this problem? Apart from the fact that it 
is easy to state, several answers can be given. One answer is that MUBs are 
useful in quantum state tomography \cite{Wootters1, Appleby}. Another has to 
do with various cryptographic protocols \cite{Cerf}. Thus, whether one wants 
to find or hide information, unbiasedness is a useful property. A third answer 
is that when one  begins to look into it, the MUB problem leads 
one into many corners of mathematics that have been explored in communication 
theory, computer science, and so on, but which are relatively unknown 
to quantum physicists. If the essence of quantum mechanics is that it 
permits one to {\it do} things that cannot be done in a classical world, 
then many surprises may be lurking in those corners.

As a matter of fact, over the past three years or so, papers about the 
MUB problem have appeared at a rate larger than once a month. 
I found about forty of those papers interesting, but I decided to quote 
only a small fraction here. On the other hand I will borrow results freely 
from people that I mention in the acknowledgements; we recently wrote a 
paper containing much more detail than did my talk (details that this 
written version will occasionally hint at) \cite{mub64}.

\vspace{1cm}

{\bf 2. Existing constructions}

\vspace{5mm}

\noindent Let us begin by taking a look at existing constructions. Recall 
that a discussion of position and momentum usually begins with the equation 

\begin{equation} [x,p] = i \ , \end{equation} 

\noindent or in terms of unitary operators 

\begin{equation} e^{iu x}e^{iv p} = e^{-iuv}e^{iv p} 
e^{iu x} \ . \end{equation}

\noindent In his 1928 book, Hermann Weyl \cite{Weyl} considered a finite dimensional 
analogue of this. We look for two unitary operators $X$ and $Z$---the 
notation is supposed to suggest an analogy to the Pauli matrices---such 
that 

\begin{equation} XZ = qZX \ , \label{Weyl} \end{equation}

\noindent where $q$ is a phase factor. Weyl found that if $q$ is a primitive 
root of unity, say if 

\begin{equation} q = e^{\frac{2\pi i}{N}} \ , \end{equation}

\noindent then eq. (\ref{Weyl}) admits a representation that is unique up 
to unitary equivalence, and the eigenbases of $X$ and $Z$ are indeed 
mutually unbiased (although this piece of terminology came later!). If the 
eigenbasis of $Z$ is chosen to be the standard basis, then the eigenbasis 
of $X$ consists of the columns of the Fourier matrix $F$, whose matrix 
elements are 

\begin{equation} F_{ab} = q^{ab} \ , \hspace{6mm} 0 \leq a, b \leq N-1 \ . 
\end{equation}

\noindent It is called the Fourier matrix because it appears in the 
discrete Fourier transform. So far, all statements are independent of 
the dimension $N$. Closer examination of the group that is generated 
by $X$ and $Z$ reveals some dimension dependent things; notably if 
$N = p =$ a prime number, then one can use the Weyl group to generate 
a set of $N+1$ MUBs. But sometimes only three MUBs turn up in this way. The special 
status of prime numbers has to do with the fact that Weyl's representation 
theorem requires a phase factor that is a primitive root of unity, that is 
a root of unity such that $q^k \neq 1$ for $k < N$. This is true for all the 
roots of unity only if $N$ is prime.   

In general the following result has been established. Let $N = p_1^{n_1}\cdot p_2^{n_2} 
\cdot \dots \cdot p_k^{n_k}$ be the prime number decomposition of $N$, with 
$p_1^{n_1} < p_2^{n_2} < \dots < p_k^{n_k}$. Then the number of constructed MUBs 
obeys 

\begin{equation} p_1^{n_1} + 1 \leq \# \ \mbox{MUBs} \leq N + 1 \ . \end{equation} 

\noindent A complete set of $N+1$ MUBs has been constructed for all 
prime power dimensions. Without going into any details (see Bandyopadhyay et al. 
\cite{Boykin} for a useful account), let me observe that 
the constructions differ somewhat depending on whether $N$ is a power of an 
even or an odd prime. In particular the behaviour of the MUBs under complex 
conjugation (relative to the standard basis) is strikingly different. I do 
not know what this means, if anything. 

When $N$ is not a prime power, the known bounds are not very sharp. 
For some very special choices of $N$ a somewhat higher lower bound is known 
\cite{Beth}, but in general the problem is wide open. For $N = 6$ several attempts to 
construct more than 3 MUBs have been made, using either group theoretical 
tricks or computer searches. There seems to be a growing consensus that 3 is the 
best one can do. Perhaps this is related \cite{Zauner,Rosu} to the fact 
that no Graeco-Latin square of order $6$ exists \cite{Fisher}. A set of $N-1$ 
Latin squares that are mutually Graeco-Latin can be used to construct a finite 
affine plane. I leave this obscure remark as a hint that there are interesting 
connections between MUBs and combinatorics; for now let me just say 
that the available evidence concerning MUBs for $N = 6$ strikes me as rather weak. 

The MUB problem appears also in other branches of learning. Indeed the same, or 
nearly the same, problem has occurred in the theory of radar signals 
\cite{Alltop}, to operator algebra theorists \cite{Popa}, in Lie algebra 
theory \cite{Kostrikin}, and in coding theory \cite{Calderbank}. Unfortunately 
all technical terms differ between these groups of authors, so it is quite 
difficult for a quantum theorist to extract 
the information gathered in any of the other fields. Suffice it to say 
that the Lie algebra theorists have the best name for the problem. They 
call it the ``Winnie--the--Pooh problem'', for reasons that would take us 
too far afield. (It has to do with a free Russian translation of a verse 
hummed by Pooh.) The actual results achieved in the various fields appear 
to be roughly the same, as far as I can tell. 

\vspace{1cm}

{\bf 3. A packing problem}

\vspace{5mm}

\noindent How should we look at the MUB problem? A first way is to view 
it as a packing problem in Hilbert space, or 
more accurately in complex projective space.  

In Hilbert space a basis can be represented as the columns 
of a unitary matrix. Assuming one basis is represented by the unit 
matrix, all bases that are MUB with respect to it must be represented 
by unitary matrices of the form (for $N = 3$, say)

\begin{equation} U = \frac{1}{\sqrt{N}}\left[ \begin{array}{ccc} 1 & 1 & 1 \\ 
e^{i\phi_{10}} & e^{i\phi_{11}} & e^{i\phi_{12}} \\
e^{i\phi_{20}} & e^{i\phi_{21}} & e^{i\phi_{22}} \end{array} 
\right] \ . \end{equation}

\noindent This is (except for the normalizing factor) a complex Hadamard 
matrix. The first row has been choosen to contain only ones by 
convention. All the vectors in all bases that are MUB with respect to 
the standard basis can therefore be found on a torus parametrized by 
$N-1$ phases. This torus has a natural interpretation as the maximal flat 
torus in complex projective space (equipped with the Fubini--Study metric). 
Now I seem to be saying that finding the MUBs is equivalent to a packing 
problem on a flat torus, but unfortunately this is not quite true, 
because the tori in complex projective space are not totally geodesic. 
What this means is that intrinsic distance on the torus does not directly 
reflect the Fubini-Study distance \cite{bok}. We do have a packing problem 
in complex projective space, but packing problems are difficult, and moreover 
their solutions tend to depend on dimension in peculiar ways. 

A straightforward approach, while we remain in the $N$ dimensional Hilbert 
space, is to begin by asking for a classification of all complex Hadamard 
matrices. But here the existing results are very incomplete 
\cite{Haagerup,Dita,Karol}. 
For any $N$, the Fourier matrix exists, hence one basis 
that is MUB relative to the standard basis always exists. Are there more? 
We can multiply the Fourier matrix from the left with a diagonal 
unitary matrix, that is to say we can multiply the rows with suitable 
phase factors, and hope that with appropriate choices of the phase factors 
we can find further bases that are MUB relative to both the standard basis 
and to the Fourier basis. We can also ask if there are further Hadamard 
matrices, not related to the Fourier matrix in this way. One would naively 
expect that the MUB problem should become easier the more such Hadamards 
one finds. This expectation is not borne out however.  

To classify (complex) Hadamard matrices, it is helpful to begin by declaring 
two Hadamard matrices to be equivalent if one can 
be reached from the other by permutations of rows and columns, and by 
multiplying rows and columns with arbitrary phase factors. Since we think of 
the columns of an Hadamard matrix as representing a basis in Hilbert space, 
these operations applied to columns mean nothing at all to us, while the same 
operations applied to rows are achieved by unitary transformations that leave 
the standard basis invariant. I will refer to the standard basis as the 
``zeroth MUB''. Then I do not loose any generality 
if I assume that the first vector in the ``first MUB'' has entries 
$1/\sqrt{N}$ only. Thus $(\mbox{MUB})_1$ is represented by an Hadamard matrix 
in dephased form (the first row and the first column has all entries real and 
positive), while any further MUBs are represented by enphased Hadamard matrices.  

When $N = 2$ or $3$ the Fourier matrix is unique up to equivalences. Hence, once 
$(\mbox{MUB})_0$ and the first vector of $(\mbox{MUB})_1$ are chosen, everything 
else is forced. But when $N = 4$ there exists a one parameter family of 
inequivalent Hadamard matrices, found (appropriately) by Hadamard himself. It is 

\begin{equation} H_4(\phi) = \frac{1}{2}\left[ \begin{array}{cccc} 1 & 1 & 1 & 1 \\ 
1 & e^{i\phi} & - 1 & - e^{i\phi} \\ 1 & - 1 & 1 & - 1 \\ 
1 & - e^{i\phi} & - 1 & e^{i\phi} \end{array} \right] \ . \end{equation}

\noindent Hence there is some freedom in choosing $(\mbox{MUB})_1$. And the 
choice matters---we have to set $\phi$ equal to zero, or to $\pi$, if we want 
to have an additional three MUBs. 

When $N = 5$ the Fourier matrix is again unique \cite{Haagerup}, but when 
$N = 6$ there are many choices. I will discuss them later. When $N$ is 
a prime number one cannot introduce any free parameters into the Fourier 
matrix, but at least when $N = 7$ another, unrelated, one parameter family 
is known. In brief, the situation is confusing, 
but based on the information available one would be inclined to guess that 
finding MUBs should be particularly easy when $N = 6$. Which is definitely 
not the case!

\vspace{1cm}

{\bf 4. The shape of the body of density matrices}

\vspace{5mm}

\noindent A second way to look at MUBs is to look at density matrices. Each 
vector in an $N$ dimensional Hilbert space can be thought of as a rank one 
projector in the set of Hermitian matrices of unit trace, which has real 
dimension $N^2 - 1$. It is convenient to think of this space as a vector 
space, with the origin placed at 

\begin{equation} \rho_\star = \frac{1}{N}{\bf 1} \ . \end{equation}

\noindent Then a unit vector $|e\rangle $ in Hilbert space corresponds 
to a real unit vector ${\bf e}$ in an $N^2 - 1$ dimensional vector space, 
whose elements are traceless matrices. The explicit correspondence is 

\begin{equation} |e\rangle \hspace{5mm} \rightarrow \hspace{5mm} {\bf e} = 
\sqrt{\frac{2N}{N-1}}\left( |e\rangle \langle e| - \rho_\star \right) 
\ . \end{equation}

\noindent Our vector space is equipped with the scalar product 

\begin{equation} {\bf e}\cdot {\bf f} = \frac{1}{2}\mbox{Tr}{\bf e}
{\bf f} \ . \end{equation}

\noindent The catch is that only a small subset of all unit vectors in the 
large vector space arises in this way. Anyway, the body of density matrices 
is now obtained by taking the convex cover of all those points on the unit 
sphere where some vector ${\bf e}$ ends. Equivalently, the body of density 
matrices consists of all Hermitian matrices of unit trace and positive 
spectrum. Either way, it is a body with an intricate shape that is difficult 
to visualize. It touches its $N^2 - 2$ dimensional outsphere in a small 
$2(N-1)$ dimensional subset, arising from vectors in Hilbert space through 
the above correspondence. These are the pure states. The whole space is 
naturally an Euclidean space, with a squared distance given by 

\begin{equation} D^2(A,B) = \frac{1}{2}\mbox{Tr}(A-B)^2 \ . \end{equation}

\noindent This is the Hilbert-Schmidt distance. The geometry induced on the 
set of pure states is of course precisely the Fubini-Study 
geometry on complex projective $(N-1)$-space. Alternatively, the Hilbert-Schmidt 
distance provides the chordal distance between two points on the outsphere of 
the body of density matrices \cite{bok}. 

An orthonormal basis in Hilbert space is now represented as a regular simplex 
with $N$ corners, spanning an $(N-1)$ dimensional flat subspace through the 
center of the outsphere. Two bases will be MUB if and only if their respective 
$(N-1)$-planes are totally orthogonal. Since the whole space has dimension 

\begin{equation} N^2 - 1 = (N+1)(N-1) \ , \end{equation}

\noindent it is clear that at most $N+1$ MUBs can be found.    

After a moment's reflection one sees that a complete set of MUBs defines a 
rather interesting convex polytope, which is the convex cover of $N+1$ regular 
simplices placed in totally orthogonal $(N-1)$-planes. It is called 
the Complementarity Polytope \cite{IB}. Evidently such a 
polytope will exist in every $N^2 - 1$ dimensional flat space. This does not 
yet solve our problem though, because given such a polytope its corners will 
correspond to vectors in Hilbert space if and only if we can rotate it so 
that all its corners fit into a special $2(N-1)$ dimensional subset of its 
$N^2 - 2$ dimensional outsphere. And when $N > 2$ this is hard!

Nevertheless this is an appealing picture of MUBs. In particular, this is the 
way to see why MUBs solve the problem of optimal state determination in 
non-adaptive quantum state tomography \cite{Ivanovic, Wootters2}. 

\vspace{1cm}

{\bf 5. Real Hilbert spaces}

\vspace{5mm}

\noindent It is 
instructive to pause to think about MUBs in real Hilbert spaces, because 
in this case it is easy to derive some negative results. Indeed real Hadamard 
matrices can exist only if $N$ is two or divisible by four. So we see 
at once that in a three dimensional real Hilbert space, it is impossible to 
find even a pair of MUBs. We can see why geometrically---in real Hilbert space, 
not among the density matrices. If a vector is 
represented by a pair of antipodal points on the unit sphere in ${\bf R}^3$, 
then a basis is represented by the corners of an octahedron. It is then 
geometrically evident that there are four vectors that are unbiased with 
respect to a given basis, and they form the corners of a cube that is dual 
to the octahedron. But they do not form a basis!

In ${\bf R}^4$ the situation is different. It is still true that a basis is 
represented by the corners of a cross polytope (the generalisation to arbitrary 
dimension of the octahedron), and there will be eight vectors that are MUB with 
respect to a given basis, again forming a cube that is dual with respect to 
the cross polytope. But specifically in four dimensions, a cube can be regarded 
as the convex cover of two symmetrically placed cross polytopes \cite{Coxeter}. 
In this way we end up with three bases represented by three symmetrically 
placed cross polytopes, and their convex cover is a famous Platonic body known 
as the 24-cell (having no analogue in three dimensions). This gives us three MUBs, 
and since the dimension of the set of real four-by-four density matrices is nine, 
this is a complete set in four dimensions. It is a set that has acquired some 
fame in quantum foundations, because a pair of dual 24-cells correspond to 24 
vectors that are uncolourable in the Kochen--Specker sense \cite{Peres}.  

If you want to know what happens in higher real dimensions, consult Boykin 
et al. \cite{BSTW}. 

\vspace{1cm}

{\bf 6. Mubness}

\vspace{5mm}

\noindent Let us 
return to the complex case. I will describe an attempt to investigate 
what happens in six dimensions, but in order to say something more interesting 
than the obvious ``I failed'', I need a measure of how much I fail. Many such 
measures, of varying degrees of sophistication, can be imagined. The one we 
use is based on the picture of MUBs that emerged from the density matrix 
point of view, namely that they correspond to totally orthogonal $(N-1)$-planes 
in an $N^2 - 1$ dimensional space. Just as vectors in an $N$ dimensional space 
can be regarded as rank one projectors in a higher dimensional space, so one 
can regard $n$-planes in an $m$-dimensional space as rank $n$ projectors in a 
vector space of sufficiently high dimension. In mathematical terms, this 
provides an embedding of the Grassmannian of $n$-planes into a flat vector 
space. The rank $n$ projectors will sit on a sphere in this flat space, and 
its natural Euclidean distance provides us with a chordal distance between 
the projectors. This notion of distance has been used to study packing problems 
for $n$-planes \cite{Conway}, and it is the one we use to measure the distance 
between bases in Hilbert space. The chordal distance attains its maximum if 
the $n$-planes are totally orthogonal, that is to say, if the bases are MUB. In 
this way it does provide a measure of ``mubness''.   

The details, adapted to our case, are as follows. Starting from a basis in 
Hilbert space, form the $N$ vectors ${\bf e}_a$. Then form the $(N^2 - 1)\times 
N$ matrix  

\begin{equation} B = \sqrt{\frac{N-1}{N}}
[ {\bf e}_1 \ {\bf e}_2 \ \dots \ {\bf e}_N] 
\ . \end{equation}

\noindent It has rank $N-1$. Next we introduce an $(N^2 -1)\times (N^2 -1)$ 
matrix of fixed trace, which is a projector onto the $(N-1)$ dimensional 
plane spanned by the ${\bf e}_a$:  

\begin{equation} P = B \ B^{\rm T} = \frac{N-1}{N}[ {\bf e}_1 
\ \dots \ {\bf e}_N]\left[ \begin{array}{c} {\bf e}^{\rm T}_1 \\ 
\dots \\ {\bf e}_N^{\rm T}\end{array} \right] \ . \end{equation}

\noindent It is easy to check (through acting on ${\bf e}_a$ say) that 
this really is a projector. The chordal Grassmannian distance between 
two $N-1$ planes spanned by two different bases then becomes 

\begin{equation} D^2_c(P_1, P_2) = \frac{1}{2}\mbox{Tr}(P_1 - P_2)^2 = 
N-1 - \mbox{Tr}P_1P_2 \ . \end{equation}

\noindent It should be obvious that there is an analogy to how the density 
matrices were defined in the first place, and to the Hilbert-Schmidt distance 
between them. The difference is that now the projectors represent entire 
bases in Hilbert space, not single vectors.

Working through the details, one finds that 

\begin{equation} D_c^2(P_1, P_2) = N - 1 - \sum_a\sum_b\left( 
|\langle e_a|f_b\rangle |^2 - \frac{1}{N}\right)^2 \ . \end{equation}

\noindent Thus 

\begin{equation} 0 \leq D^2_c \leq N-1 \ , \end{equation} 

\noindent and the distance attains its maximum value if and only if the 
bases are MUB. As a measure of mubness, the chordal Grassmannian distance 
has the advantages that it is geometrically natural and simple to compute. 
It is also natural from the tomographic point of view, although I certainly 
cannot claim any precise operational meaning for it.
 
\vspace{1cm}

{\bf 7. N = 6}

\vspace{5mm}

\noindent My third and final way to look at MUBs is to simply perform calculations to 
see what happens, without thinking very much. We tried the first open case: 
$N = 6$. Maybe this was a mistake, because $6$ clearly sits astride the even 
and the odd prime numbers. Perhaps we should concentrate on $15 = 3\cdot 5$? 
(The question how many mutually Graeco-Latin squares exist is actually open 
in this case.) On the other hand, a six dimensional Hilbert 
space is already a very large space to search in---and fifteen is larger. So 
we stick to six.

Let $(\mbox{MUB})_0$ be the standard basis, and $(\mbox{MUB})_1$ be given 
by the columns of some dephased Hadamard matrix (that is one whose first row 
and first column are real). We 
make a choice for $(\mbox{MUB})_1$, find all enphased Hadamard matrices that 
represent bases that are MUB with respect to $(\mbox{MUB})_1$, and then 
check how far apart the latter are, in the sense of the chordal distance.

When $N = 6$ there are several choices for $(\mbox{MUB})_1$. The following 
dephased Hadamard matrices are known \cite{Karol}:

\

\noindent i) The Fourier matrix, augmented with two free parameters:

\begin{equation} F_6(\phi_1,\phi_2) = 
\left[ \begin{array}{cccccc} 1 & 1 & 1 & 1 & 1 & 1 \\ 
1 & q^{i\phi_1} & q^2e^{i\phi_2} & q^3 & 1^4e^{i\phi_1}& q^5e^{i\phi_2} \\
1 & q^2 & q^4 & 1 & q^2 & q^4 \\
1 & q^3e^{i\phi_1} & e^{i\phi_2} & q^3 & e^{i\phi_1} & q^3e^{i\phi_2} \\
1 & q^4 & q^2 & 1 & q^4 & q^2 \\
1 & q^5e^{i\phi_1} & q^4e^{i\phi_2} & q^3 & q^2e^{i\phi_1} & qe^{i\phi_2} 
\end{array}\right] , \hspace{5mm} q \equiv e^{\frac{2\pi i}{6}} \ . \end{equation}

\noindent There are various equivalences of the 
form $F(\phi_1, \phi_2) \approx F(\phi_3, \phi_4)$; this has been sorted out, but 
this is not the place to give all the details.  

\noindent ii) The transpose of the above, again with two free parameters.

\noindent iii) A one-parameter family of matrices $D(\phi)$ whose entries, when 
the free phase is set to zero, are fourth roots of unity. It is known as the 
Di\c{t}\u{a} family.

\noindent iv) An isolated matrix $S$ whose entries are third roots of unity.

\noindent v) A matrix found by Bj\"orck, which is most conveniently given as the 
circulant matrix
 
\begin{equation} C = \left[ \begin{array}{cccccc} 
1 & id & - d & - i & - \bar{d} & i\bar{d} \\
i\bar{d} & 1 & id & - d & - i & - \bar{d} \\
- \bar{d} & i\bar{d} & 1 & id & - d & - i \\
- i & - \bar{d} & i\bar{d} & 1 & id & - d \\
- d & - i & - \bar{d} & i\bar{d} & 1 & id \\
id & - d & - i & - \bar{d} & i\bar{d} & 1 \end{array} \right] \ . 
\end{equation}

\noindent A matrix is said to be circulant if its columns are cyclic permutations 
of its first column. The complex number $d$ has modulus unity and is 

\begin{equation} d = \frac{1-\sqrt{3}}{2} + i\sqrt{\frac{\sqrt{3}}{2}} 
\hspace{5mm} \Rightarrow \hspace{5mm} 
d^2 - (1-\sqrt{3})d + 1 = 0 \ . \end{equation}

\

\noindent Several people have invested some effort in making this list as 
long as it can be, and up to two weeks before my talk, I thought that it 
might well be a complete list. Still, we do have considerable latitude in 
how we choose the first MUB. 

Let us begin with 

\begin{equation} (\mbox{MUB})_1 = F(0,0) \ . \end{equation}

\noindent It happens that all vectors that are unbiased with respect to 
this choice of the zeroth and first MUB have been computed, first by Bj\"orck 
and Fr\"oberg \cite{Bjorck}, and independently by Grassl \cite{Grassl}. 
Bj\"orck did not express the problem in these terms however. He was interested 
in biunimodular sequences, that is to say sequences of unimodular complex 
numbers $x_a$ whose discrete Fourier transform 

\begin{equation} \tilde{x}_a = \frac{1}{\sqrt{N}}\sum_{a = 0}^{N-1} x_bq^{ba}
\end{equation}

\noindent also consists of unimodular complex numbers. On reflection, one 
sees that the two problems are equivalent. Bj\"orck and coauthors eventually 
solved this problem for all $N \leq 8$. When $N = 6$ there are altogether 48 
such sequences; 12 Gaussian ones---they were known to Gauss---and an 
additional 36. The Gaussian ones have entries that are 12th roots of unity, 
while the additional ones involve Bj\"orck's magical number $d$. See also 
Haagerup \cite{Haagerup}, who seems to have been the first to get this 
quite right. 

More is true. For a biunimodular sequence the autocorrelation function is    

\begin{equation} \gamma_b \equiv \frac{1}{N}\sum_{a=0}^{N-1}\bar{x}_a
x_{a+b} = \frac{1}{N}\sum_{a=0}^{N-1}
|\tilde{x}_a|^2q^{-ab} = \delta_{b,0} \ . \end{equation} 

\noindent Therefore $x_a$ and $x_{a+b}$, with $b$ fixed and non-zero, 
are orthogonal vectors. Then it follows that the 12 + 36 vectors found by 
Bj\"orck can be assembled into 2 + 6 unitary circulant matrices that are 
MUB with respect to the standard basis and the Fourier basis. When Grassl 
redid this calculation (using the program MAGMA) he observed that each of 
the 48 vectors can be used to form a basis in exactly two ways. Thus we 
end up with exactly 2 + 2 + 6 + 6 = 16 possible choices for $(\mbox{MUB})_2$. 
In itself, this is more than we need for a complete set of 7 MUBs. Provided 
that the Fourier basis is included, the question whether one can find a fourth 
MUB boils down to the question whether the chordal distance squared between 
any pair among the 16 is equal to 5, the maximal distance squared attained by 
MUBs in six dimensions. 

The answer is no. The 4 Gaussian MUBs, composed of 12th roots, form a 
perfect square with side lengths squared $D_c^2 = 2$. The two groups of 
``non-Gaussian'' MUBs are isometric copies of each other. One group consists 
of circulant matrices, while the other group consists of Fourier matrices 
enphased using the magical number $d$. The distance squared between any 
Gaussian and any non-Gaussian MUB is $D_c^2 \approx 4.62$ ---rather close 
to $5$, if one takes an optimistic view of things. The distance between the 
two six-plets of non-Gaussian MUBs is always $D^2_c \approx 3.71$. Inside 
each group, distances reach all the way up to $D^2_c \approx 4.64$, which 
is even closer to $5$. 

Still, although the pattern is nice, the conclusion is negative: the Fourier 
matrix cannot be included in a set of more than three MUBs. We do not have 
complete results for other choices of the first MUB. We made a program that 
lists all MUB triplets where all the entries of the matrices are 24th roots of 
unity. Quite a few triplets, with quite interesting structures, did turn 
up in this way, but there were no MUB quartets. 
Note that in prime power dimensions the standard solution for complete sets 
of MUBs contain $N$th or $2N$th roots of unity only (depending on whether 
$N$ is odd or even \cite{Wootters2}); in other words the analogous calculation 
in arbitrary dimension would have found the known complete sets.

Although the evidence is incomplete, we do seem to be driven to admit that 
there can exist at most 3 MUBs when $N = 6$. A nagging doubt remains, 
because there is always the possibility that the above list of Hadamard 
matrices is incomplete. In particular, could the parameter spaces be incomplete? 
We do know that the number of free parameters in Fourier's, Di\c{t}\u{a}'s, 
and Bj\"orck's matrices is at most 4, while the matrix $S$ cannot have any 
free parameters at all \cite{Tadej}. But we do not know if there are that 
many free parameters.   
   
While I was thinking about what to say in my talk, Beauchamp and Nicoara 
\cite{Nicoara} intervened. They found a new one-parameter family of Hadamard 
matrices, namely 

\begin{equation} B = \left[ \begin{array}{cccccc} 
1 & 1 & 1 & 1 & 1 & 1 \\
1 & - 1 & - \frac{1}{x} & - y & y & \frac{1}{x} \\
1 & -x & 1 & y & \frac{1}{z} & - \frac{1}{t} \\
1 & - \frac{1}{y} & \frac{1}{y} & -1 & -\frac{1}{t} & \frac{1}{t} \\
1 & \frac{1}{y} & z & -t & 1 & -\frac{1}{x} \\
1 & x & -t & t & -x & -1 \end{array} \right] 
\end{equation}

\noindent where $(x,y,z,t)$ are complex numbers of modulus one, related by 

\begin{equation} t = xyz \end{equation}

\begin{equation} z = \frac{1+2y-y^2}{y(-1+2y+y^2)} \end{equation}

\begin{equation} x = \frac{1+2y +y^2 \pm \sqrt{2}\sqrt{1+2y+2y^3+y^4}}
{1+2y-y^2} \ , \end{equation}

\noindent with $y$ remaining as a free phase factor. By 
construction this family contains all $N = 6$ Hermitian Hadamard matrices. 
The phase of $y$ cannot 
be chosen quite arbitrarily; an interval around $y = 1$ is excluded. 
On closer inspection one finds that this family starts from 
a matrix that is equivalent to Bj\"orck's, passes through the Di\c{t}\u{a} 
family, comes back to Bj\"orck, repeats twice, and ends at Bj\"orck. The two 
branches of the square root lead to equivalent families. 

What does this mean? I do not know. If it means that there are four dimensional 
families of Hadamard matrices, including the Fourier matrix and Bj\"orck's matrix, 
then it also means that we have looked in a very small part of parameter space only. 
In fact we do have some reasons to believe that this is really so. Therefore it 
seems to 
me that the conclusion that we must draw about the MUB 
problem in six dimensions is: We have almost no evidence either way.  

\vspace{1cm}

{\bf 8. The real problem}

\vspace{5mm}

\noindent The distance that we introduced can, in principle, be used to convert 
the MUB problem into that of maximising a function, such as 

\begin{equation} F = \sum_{i,j}D^2_c(P_i,P_j) \ . \end{equation}

\noindent A similar procedure has been used \cite{Renes} to find approximations to 
SIC-POVMs---this particular acronym stands for a kind of relative of the 
MUB problem---in dimensions up to $N = 45$. In our case the upper bound is 
attained if the $N+1$ projectors represent totally orthogonal $(N-1)$-planes. 
Whether we can reach the upper bound using $(N-1)$-planes spanned by bases in 
the underlying Hilbert space is of course precisely the question. 
  
I should add that I have not really done justice to the point of view that I 
tried to stress in the beginning, that the MUB problem leads one into many 
corners of useful mathematics that have not been very much explored by quantum 
physicists. But if you search your favourite eprint archive for some of the 
many papers, whose existence I hinted at, you will see what I mean. Meanwhile, 
the somewhat botanical spirit of my talk is perhaps appropriate in the town 
where Linnaeus was educated. 

Anyway the real ``MUB problem'' is not how many MUBs we can find. 
The real MUB problem is to find out what we can do with those that exist.

\vspace{15mm}

{\bf Acknowledgments}

\vspace{5mm}

\noindent I thank \AA sa Ericsson, Jan-\AA ke Larsson, Wojciech Tadej, 
Wojciech Bruzda, and Karol \.Zyczkowski for collaborating with me. We 
thank Markus Grassl for sending his vectors for inspection, Bengt Nagel 
for insights, and Piotr Badziag for a good question. Three of us 
thank Andrei 
Khrennikov for inviting us to visit Sm\aa land.

\end{document}